\documentclass[aps,prl,twocolumn,twoside]{revtex4}
\usepackage{bm}
\usepackage{graphicx}
\usepackage{amsfonts}
\usepackage{amsmath}
\usepackage{amssymb}
\newcommand {\be}{\begin{equation}}
\newcommand {\ee}{\end{equation}}
\newcommand {\bea}{\begin{eqnarray}}
\newcommand {\eea}{\end{eqnarray}}

\begin{document}

\title{Coherent control near metallic nanostructures}

\author{Ilya Grigorenko}
\affiliation{Theoretical Division T-17, Center for Nonlinear
Studies, Center for Integrated Nanotechnologies, Los Alamos National
Laboratory, Los Alamos, New Mexico 87545, USA}
\author{Anatoly Efimov}
\affiliation{Center for Integrated Nanotechnologies, Los Alamos
National Laboratory, Los Alamos, New Mexico 87545, USA}

\date{\today}

\begin{abstract}
We study coherent control in the vicinity of metallic
nanostructures. Unlike in the case of control in gas or liquid
phase, the collective response of electrons in a metallic
nanostructure can significantly enhance different frequency
components of the control field. This enhancement strongly depends
on the geometry of the nanostructure and can substantially modify
the temporal profile of the local control field. The changes in the
amplitude and phase of the control field near the nanostructure are
studied using linear response theory. The inverse problem of finding
the external electromagnetic field to generate the desired local
control field is considered and solved.
\end{abstract}

\maketitle

 %\section{Optimal control in gas or liquid phase}

Last two decades have witnessed the birth and rapid progress in
manipulation of quantum systems, such as atoms and molecules
\cite{rabitz0,firstpub,gerber}, semiconductor quantum dots
\cite{bonadeo} or even complex biological systems \cite{bio} by
means of optimally shaped ultrashort laser pulses. Coherent control
is usually achieved in homogeneous gas or liquid phase by the direct
coupling of electronic or molecular degrees of freedom of the system
being controlled to the external electromagnetic fields. However, it
is known that most of the important chemical reactions in nature and
technology take place at surfaces or interfaces. For example, at the
surface of a catalyst (usually metal) breaking and formation of
chemical bonds takes place on a femtosecond time scale and
observation of the dynamics can be performed, e.g., using a
pump-probe technique \cite{pump_probe}. The induction and control of
chemical reactions, such as oxidization of CO on a metal surface,
using ultrashort laser pulses is very important from scientific and
technological points of view \cite{petek_review,review}.

For coherent control of atoms or molecules near metallic surfaces,
metallic clusters or other nanostructures the situation becomes very
different from the control in homogeneous dielectric media. The
interaction of valence electrons with external field may lead to
collective, geometry-dependent excitation modes and produce local
fields few orders of magnitude larger than the incident control
field \cite{stockman}. In this case the direct coupling of the
control field to the controlled molecule may be much weaker than
indirect coupling via interaction with the excited plasmon modes.
Note, that one has to distinguish between the static spatial
localization of the induced field near sharp edges of metallic
nanostructure and excitation of spatially localized plasmonic modes.
The interaction of electromagnetic field with a metallic
nanostructure, unlike in the case of homogeneous dielectric medium,
creates highly inhomogeneous induced local field distribution. The
spatial variations of the field intensity may be significant on a
scale as small as one nanometer \cite{stockman}. The effect of the
presence of a metallic nanostructure is not limited to the local
field enhancement only, but may also result in the broadening of a
state linewidth in a quantum system due to dynamical screening
effects. For example, recent measurements demonstrate \cite{fluer}
reduction of fluorescence lifetime of dye molecules (from $2.8$ ns
to less then $1$ ns) in the presence of nanostructured metal
surface.

Since {\it ab intio} simulations of a controlled quantum system and
electrons in the nanostructure may be computationally extremely
expensive, most of the previous modeling of temporal characteristics
of the local fields and associated quantum dynamics was typically
performed for oversimplified models, with the focus on some
particular aspects of the problem.
For example,
%a hybrid quantum
%control of photodesorption of NO molecule from a Pt(111) metal
%surface was considered theoretically \cite{no_desorption}. In this
%work the authors studied the influence of  strong dissipation due to
%coupling of the molecule to the metal surface on optimal control.
%However, the authors did not consider the collective response of
%electrons at the metal surface that would lead to the local field
%enhancement and distortion of the control. In another study,
a quantum mechanical treatment of time-dependent screening and
associated transient effects due to sudden creation of a charge in a
close vicinity of two dimensional electron gas, was considered in
\cite{surface_science}. Other simulations, involving more realistic
geometries than an idealized two-dimensional electron gas, are
usually performed using classical Finite Difference Time Domain
(FDTD) approach. One can mention coherent control of nanoscale
localization of optical excitations in nanosystems \cite{stockman},
where  classical equations with local phenomenological dielectric
constant $\epsilon({\bf r})$ were used. However, the validity of
local classical response theory is questionable for relatively small
or highly inhomogeneous nanostructures, where quantum mechanical
effects may play a critical role. In the quantum limit one needs to
take into account the discreteness of the energy spectrum, non-local
electron density response, dynamical screening effects and tunneling
\cite{levi}.

%
%Actually one can consider an optimal design problem : to find a
%geometry of a nanostructure to trap certain species of molecules at
%target frequency of the external field.

%\section{Challenge}
As can be seen, the coherent control of atoms and molecules
localized near metallic nanostructures is much more complex than in
a gas or liquid phase. It is important for the theory to capture the
fact that the incident electromagnetic field is coupled both to
single particle and collective plasma excitations in the
nanostructure, and the field in the nanostructure is neither
completely screened, nor purely transverse. To develop the theory of
the coherent control at the nano (i.e. subwavelength) scale, we need
to map out spatio-temporal or spatio-spectral amplitude as well as
the phase of the local field in relation to the external
spatially-uniform, but temporally shaped excitation field. More
importantly, the inverse problem of finding the external field
corresponding to the required local control field  for the best
performance needs to be addressed. It may seem that the modification
of the control field in the vicinity of a metal nanostructure makes
coherent control very difficult or impossible. However, we will show
that one can successfully generate local control fields with the
prescribed temporal behavior and take advantage of local field
enhancement.

In this study we consider the problem of coherent control in the
presence of a doped semiconductor nanostructure and propose to use
the local field enhancement due to geometrical and resonant plasmon
excitation for optimal control of atoms and molecules. Relatively
long relaxation and dephasing times were reported for heavily doped
semiconductor heterostructures \cite{dephasing}. The carrier
concentration in the nanostructure is chosen to generate plasmon
resonance around  $0.8$~eV, which corresponds to the wavelength
$\lambda=1550$~nm. The experimental techniques for generation,
shaping and detection of ultrashort pulses at this wavelength are
readily available. Also, this wavelength range is important for
excitation and optimal control of molecular wave packet formation
using a two-color pump-probe scheme \cite{kaiser}. The Fermi
wavelength at this carrier concentration is $\approx 10$~nm,
comparable to the size of the nanostructure (several tens of
nanometers-within capabilities of modern nanofabrication methods),
which motivates us to use quantum nonlocal linear response theory
\cite{levi}. Using the theory \cite{levi}, we examine the
inhomogeneous distribution of the induced field, and study how the
control pulse intensity and phase are transformed due to the
plasmonic response of a small custom-engineered nanostructure.

In the self-consistent treatment the induced charge density in the
nanostructure $\rho_{\text{ind}}({\bf r},t)$ is given by:
\begin{eqnarray}
\label{rho_ind} \rho_{\text{ind}}({\bf
r},t)=\int_{-\infty}^{+\infty}dt^\prime\int d{\bf r'}\chi_0({\bf
r},{\bf r'},t,t^\prime)\phi_{\text{tot}}({\bf r'},t^\prime),
\end{eqnarray}
where $\phi_{\text{tot}}({\bf r},t)=\phi_{\text{ext}}({\bf
r},t)+\phi_{\text{ind}}({\bf r},t)$ is the total potential, and
$\chi_0({\bf r},{\bf
r'},t,t')=-\theta(t-t')<\rho(r,t),\rho(r',t')>_0$ is the electron
polarizability, and the statistical average is performed over the
equilibrium state. The function $\theta(t-t')$ ensures the causality
of the electron response.

We assume the density-density response function is translation
invariant in time variables $\chi_0({\bf r}, {\bf r'},t,t')\equiv
\chi_0({\bf r}, {\bf r'},t-t')$, but not in the space variables. By
making Fourier transform in time domain and assuming Random Phase
Approximation (RPA), one can express $\chi_0({\bf r}, {\bf
r'},\omega)$ in terms of the eigenenergies and eigenfunctions of the
the unperturbed Hamiltonian:
\begin{equation}
\chi_0({\bf r}, {\bf
r'},\omega)=\sum_{i,j}\frac{f(E_i)-f(E_j)}{E_i-E_j-\hbar\omega-i\gamma}\psi_i^*({\bf
r})\psi_i({\bf r'})\psi_j^*({\bf r'})\psi_j({\bf r}),
\end{equation}
where $f$ is the Fermi distribution function, and $\gamma$ accounts
for the finite width of the quantum levels in the nanostructure.
 The eigenproblem for electrons confined in the nanostructure $\label{eq1} \label{hamiltonian} H \psi_i({\bf r})
 = E_i \psi_i({\bf r})$ is solved  using numerical diagonalization.

 The induced potential $\phi _{\text{ind}}({\bf r},\omega)$ is
determined from the self-consistent integral equation \cite{levi}
\bea \label{integral_equation} \phi _{{\text {ind}}} ({\bf
r},\omega) = \int{\int {\chi_0 ({\bf r'},{\bf r''},\omega )\phi
_{{\text {tot}}} ({\bf r''},\omega) V_C(|{\bf r}-{\bf r'}|) d{ \bf
r'}} d {\bf r''}}, \eea where $V_C(|{\bf r}-{\bf r'}|)$ is the
Coulomb potential. Here we make use of the fact that the
nanostructure is much smaller than the wavelength of the incoming
field, so the retardation effects can be neglected. In order to find
$\phi _{{\text {ind}}} ({\bf r},\omega)$ we discretize
Eq.(\ref{integral_equation}) on a real-space cubic mesh and solve it
numerically. The equation Eq.(\ref{integral_equation}) establishes
one-to-one correspondence between the homogeneous external control
field ${\bf E}_{\text{ext}}(\omega)$ and the total (local) field
${\bf E}_{\text{tot}}({\bf r},\omega)$ at a given point ${\bf r}$.
Note that this correspondence is established using statistical
average and has no uncertainties related to quantum dynamics of
electrons for nanostructures with large number of electrons.

Since we work in a weak-field limit and assume linear response of the system,
there is no mixing or creation of new frequency components, which were not
originally present in the incoming control field. This linear response
approximation significantly simplifies the analysis in that the response of the
system can be computed for each frequency separately in the spectral domain:

\begin{eqnarray}
\label{frequency2} {\bf E}_{\text{tot}}({\bf r}_0,\omega)={\bf {\hat
Z}}({\bf r}_0,\omega){\bf E}_{\text{ext}}(\omega),
\end{eqnarray}
where ${\bf r}_0$ is the position of the controlled molecule, and
for a small metallic nanostructures the complex valued local
response kernel ${\bf {\hat Z}}({\bf r}_0,\omega)$ is non-zero for
any $\omega$. Thus, the expression Eq.(\ref{frequency2}) can be
inverted, and the optimal external field can be computed given the
required local control field ${\bf E}_{\text{tot}}({\bf
r}_0,\omega)$.

% However, due to...
\begin{figure}[h]
\includegraphics[width=4.7cm,angle=0,bb=110 380 500 730]{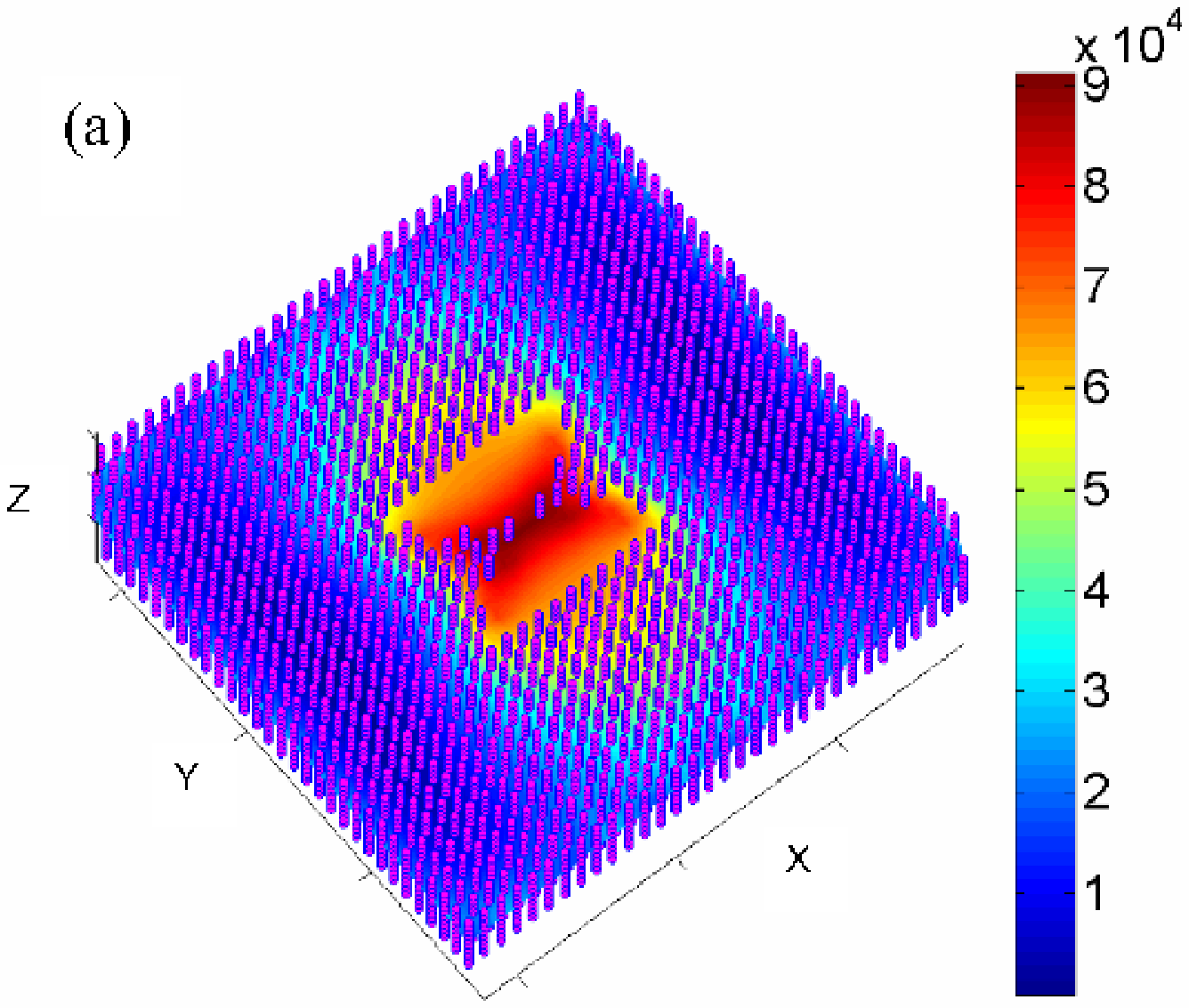}
\includegraphics[width=4.7cm,angle=0, bb= 92 238 488 540]{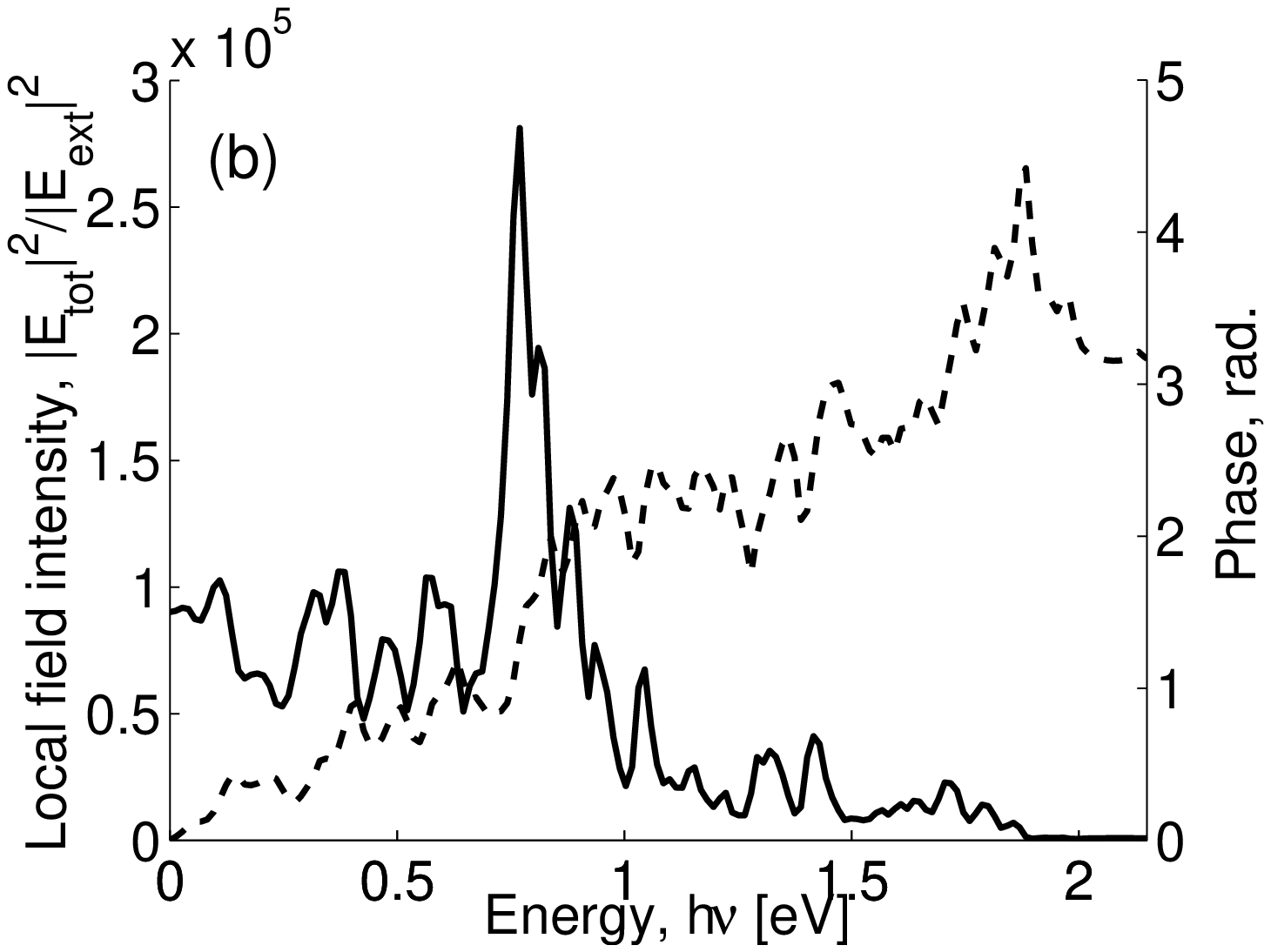}
\caption{\label{fig1} \footnotesize{(a) Normalized local field
intensity $|{\bf E}_{\text{tot}}({\bf r})|^2/|{\bf
E}_{\text{ext}}|^2$ in the nanostructure with a bow-tie shaped hole,
for static ($\omega=0$) external field. Note strong field
enhancement near the sharp edges. (b) Normalized local field
intensity $|{\bf E}_{\text{tot}}({\bf r}_0)|^2/|{\bf
E}_{\text{ext}}|^2$ (solid line) and phase (dashed line) in the
center of the nanostructure as a function of photon energy of the
external field $h\nu$. Note many local extrema due to multiple
geometric resonances in the nanostructure. The resonance peak occurs
at $h\nu\approx 0.8$ eV as the phase reaches $\pi/2$ value.}}
\end{figure}

To illustrate our approach we consider a finite nanostructure with
dimensions
$L\times L\times d $, $L=33$~nm, $d=10$~nm, %(V=1.13$\times10^{-22}$m$^3$),
made of a doped semiconductor with the carrier concentration
$n=9.5\times 10^{19}$cm$^{-3}$ with  a bow-tie shaped hole,
Fig.\ref{fig1}(a). Nanostructures of such shape are routinely
manufactured for local field enhancement \cite{bowtie}. There are
total $1042$ electrons confined in the nanostructure. The electron
effective mass is taken to be $0.2 m_e$, where $m_e$ is the bare
electron mass \cite{gamma}. The carrier concentration is chosen to
generate plasmon resonance around $0.8$~eV.

In our simulations we set the phenomenological damping parameter to
$\gamma=3.8$~meV \cite{dephasing}. We performed simulations assuming
room temperature $T=300$~K~$\approx25.8$~meV and linearly polarized
external field with the polarization direction parallel to the $x$
axis.

   Figure ~\ref{fig1}(a) shows spatial
distribution of the normalized total field intensity $|{\bf
E}_{\text{tot}}({\bf r})|^2/|{\bf E}_{\text{ext}}|^2$ in the static
($\omega\to0$) limit.  Note strong field enhancement in the vicinity
of the bow-tie sharp edges. We assume that controlled molecule is
localized in the center of the hole, in the point of local field
maximum, which is our target region.

We performed calculations of the local field ${\bf E_{\text{tot}}}$
in the target region for different frequencies of the external
field, Fig.~\ref{fig1}(b). Note that the main plasmonic resonance
peak occurs at $\omega\approx 0.8$ eV  as the phase reaches $\pi/2$
value. The frequency dependence has many local extrema due to
multiple geometric resonances in the nanostructure. Using the
spectral response data of Fig.~\ref{fig1}(b), the temporal structure
of the local field induced by an arbitrary incident field can be
uncovered, Eq.~(\ref{frequency2}), by Fourier transforming the
product of the response of the nanostructure and the incident field
complex spectrum. Similarly, the external control field in the
frequency domain can be computed simply by dividing the given local
field complex spectrum by the nanostructure response function.

\begin{figure}[h]
%\vspace{0.cm}
\includegraphics[width=4.cm,angle=0, bb= 92 238 488 540]{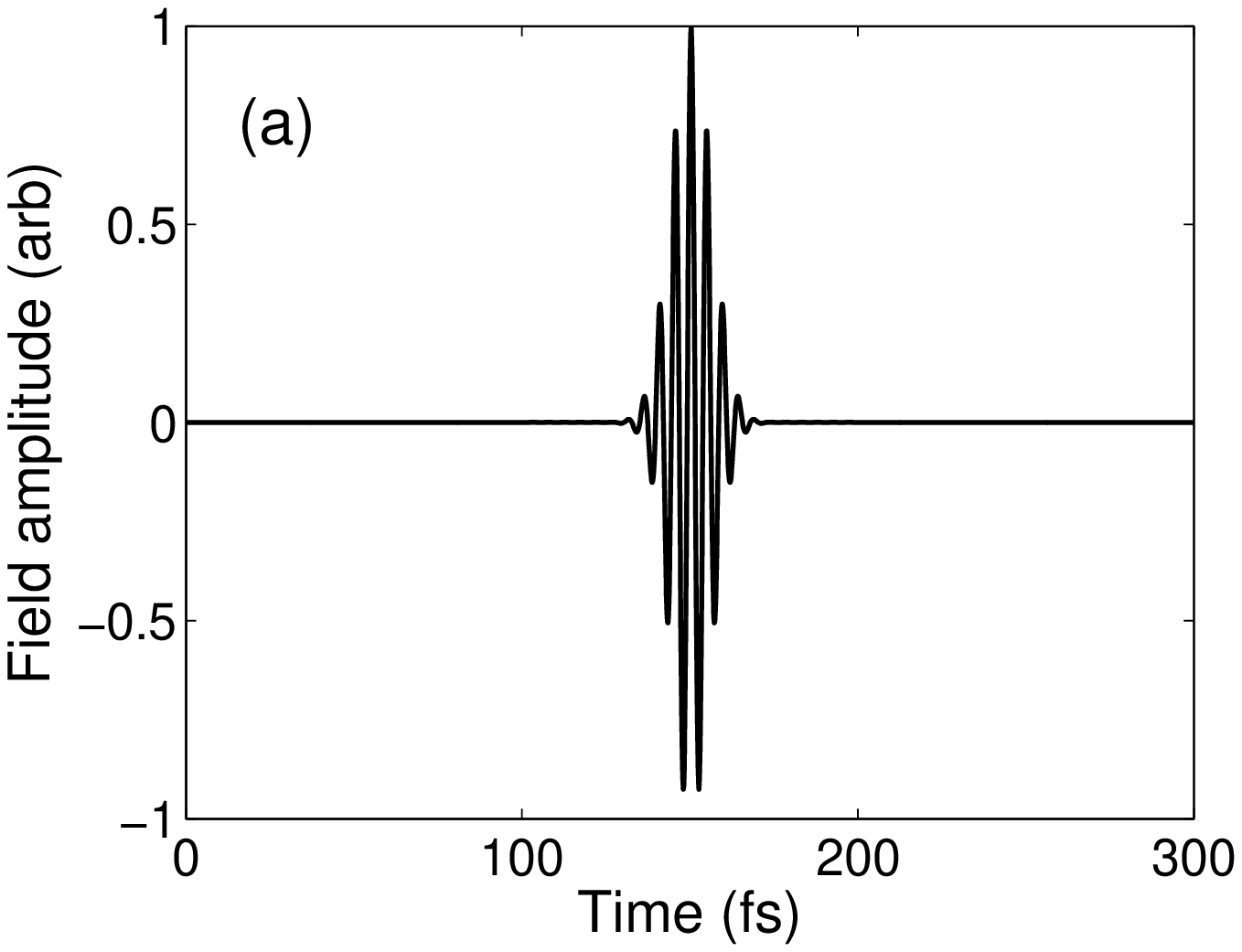}
\includegraphics[width=4.cm,angle=0, bb= 92 238 488 540]{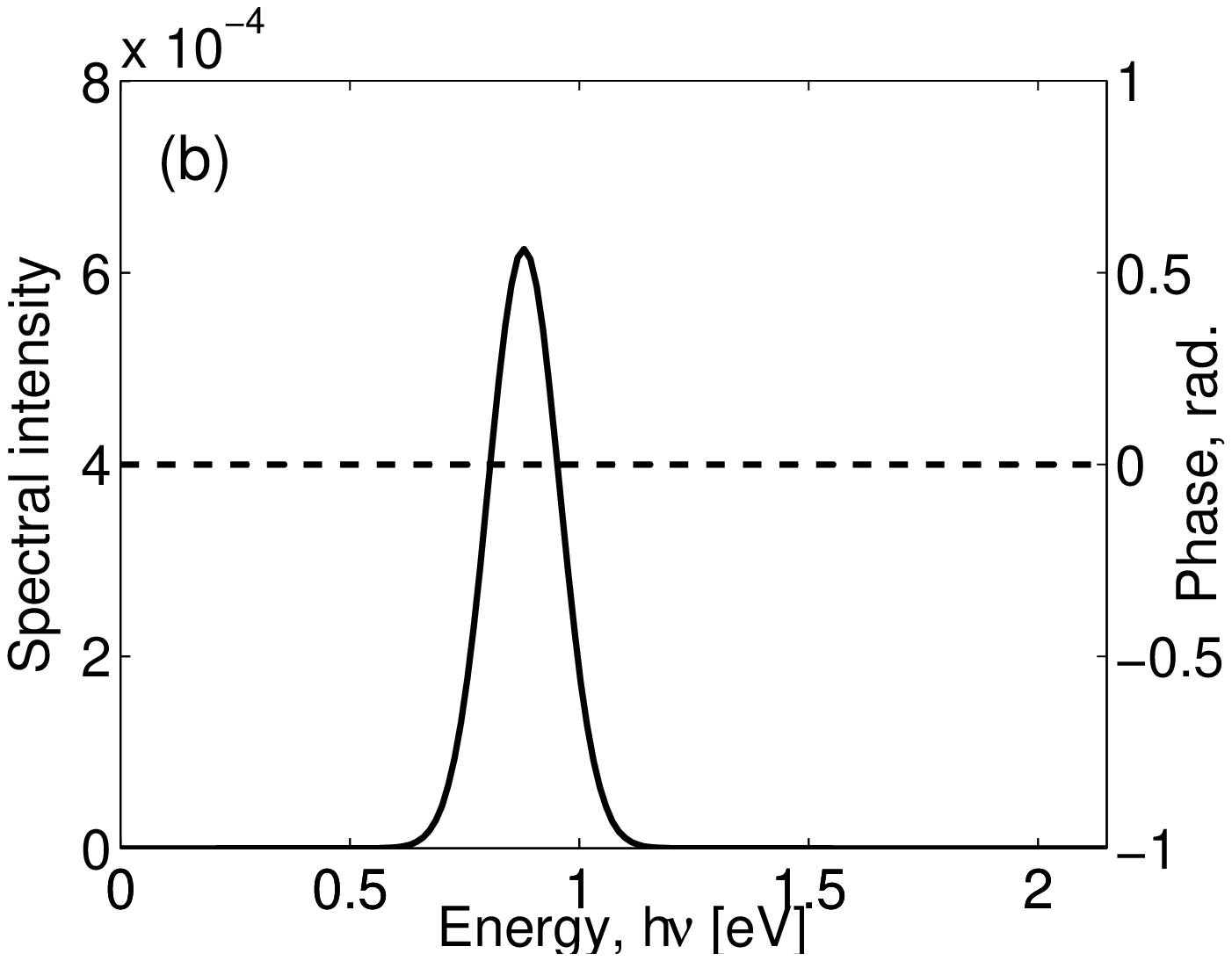}
\includegraphics[width=4.cm,angle=0, bb= 92 238 488 540]{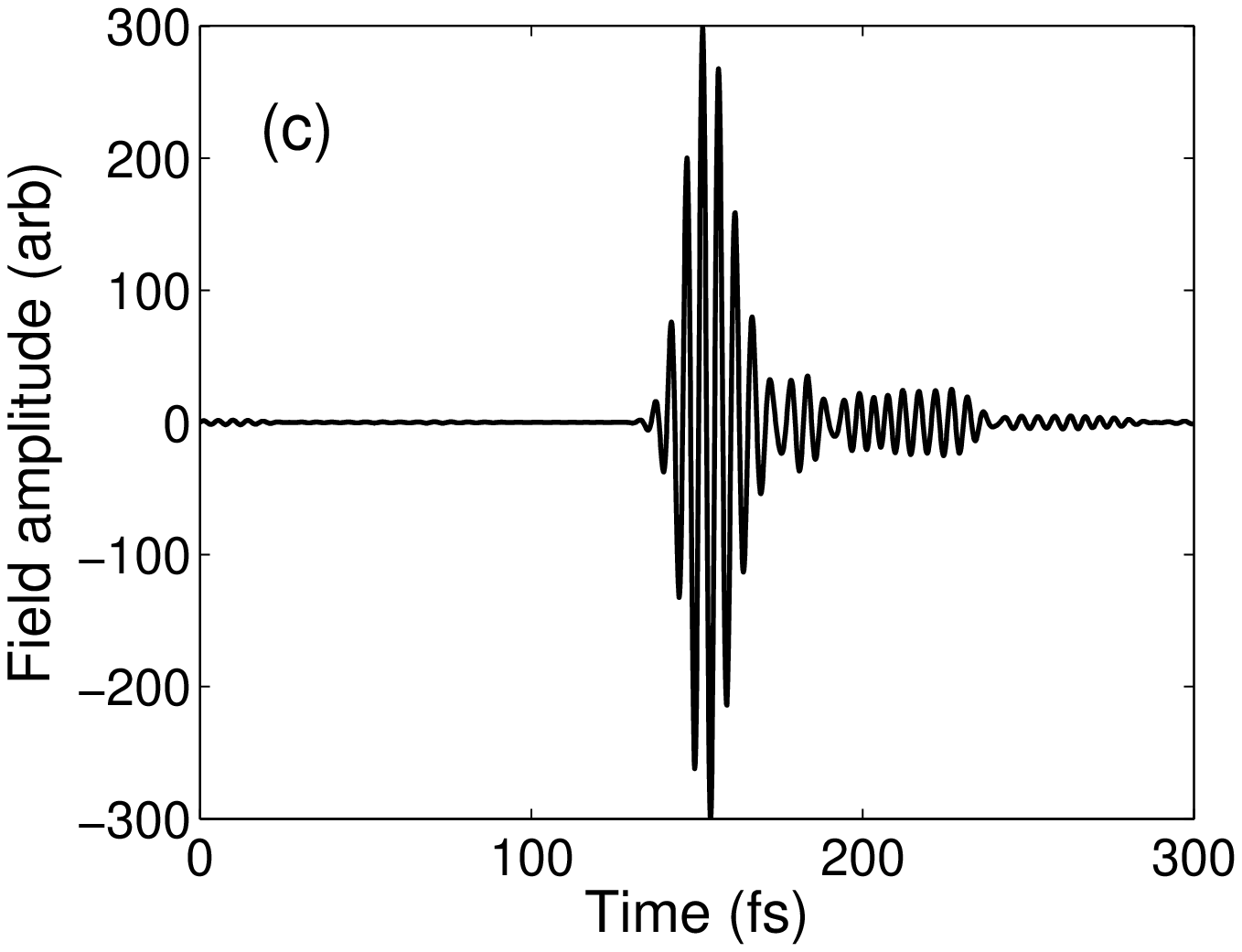}
\includegraphics[width=4.cm,angle=0, bb= 92 238 488 540]{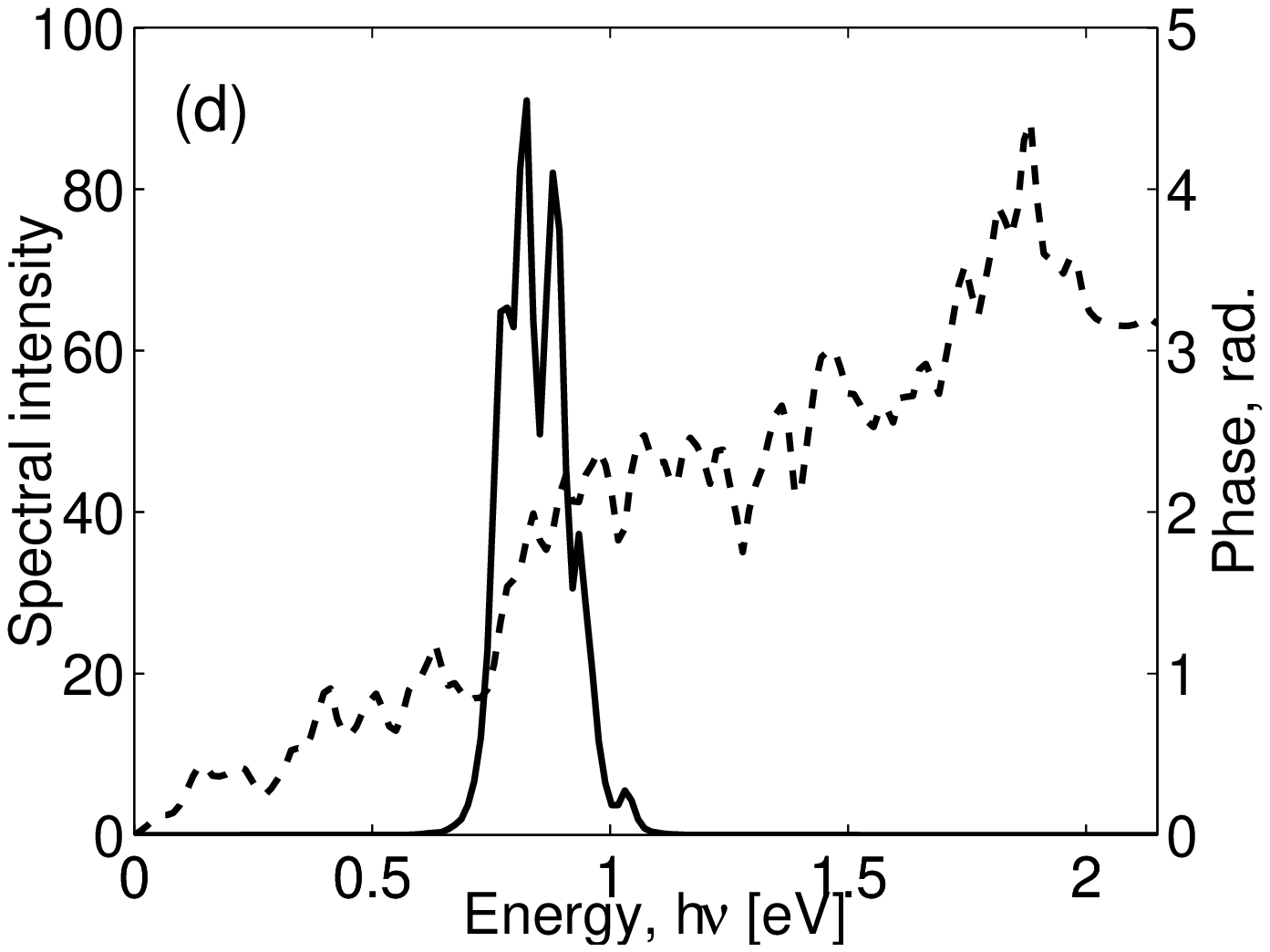}
\includegraphics[width=4.cm,angle=0, bb= 92 238 488 540]{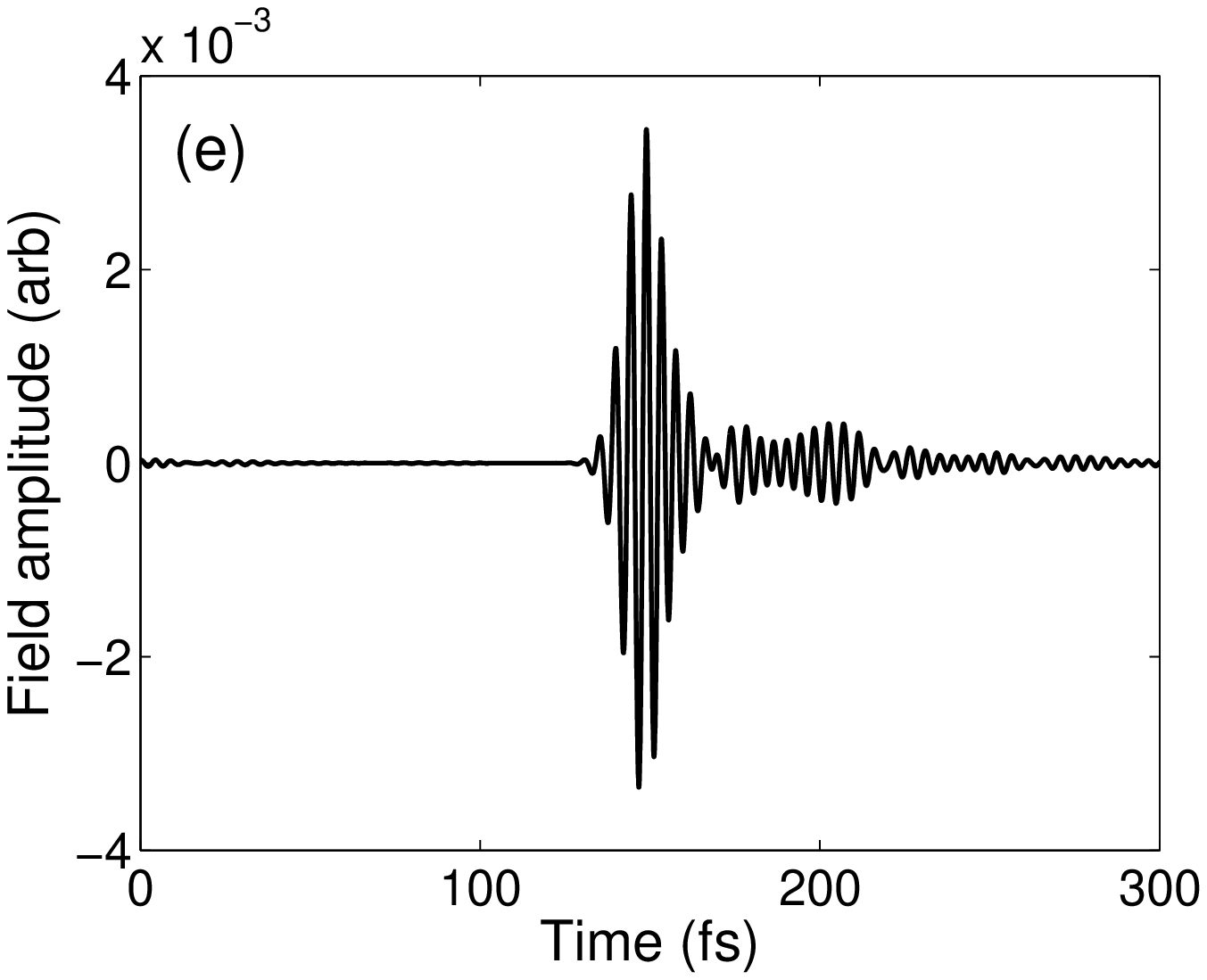}
\includegraphics[width=4.cm,angle=0, bb= 92 238 488 540]{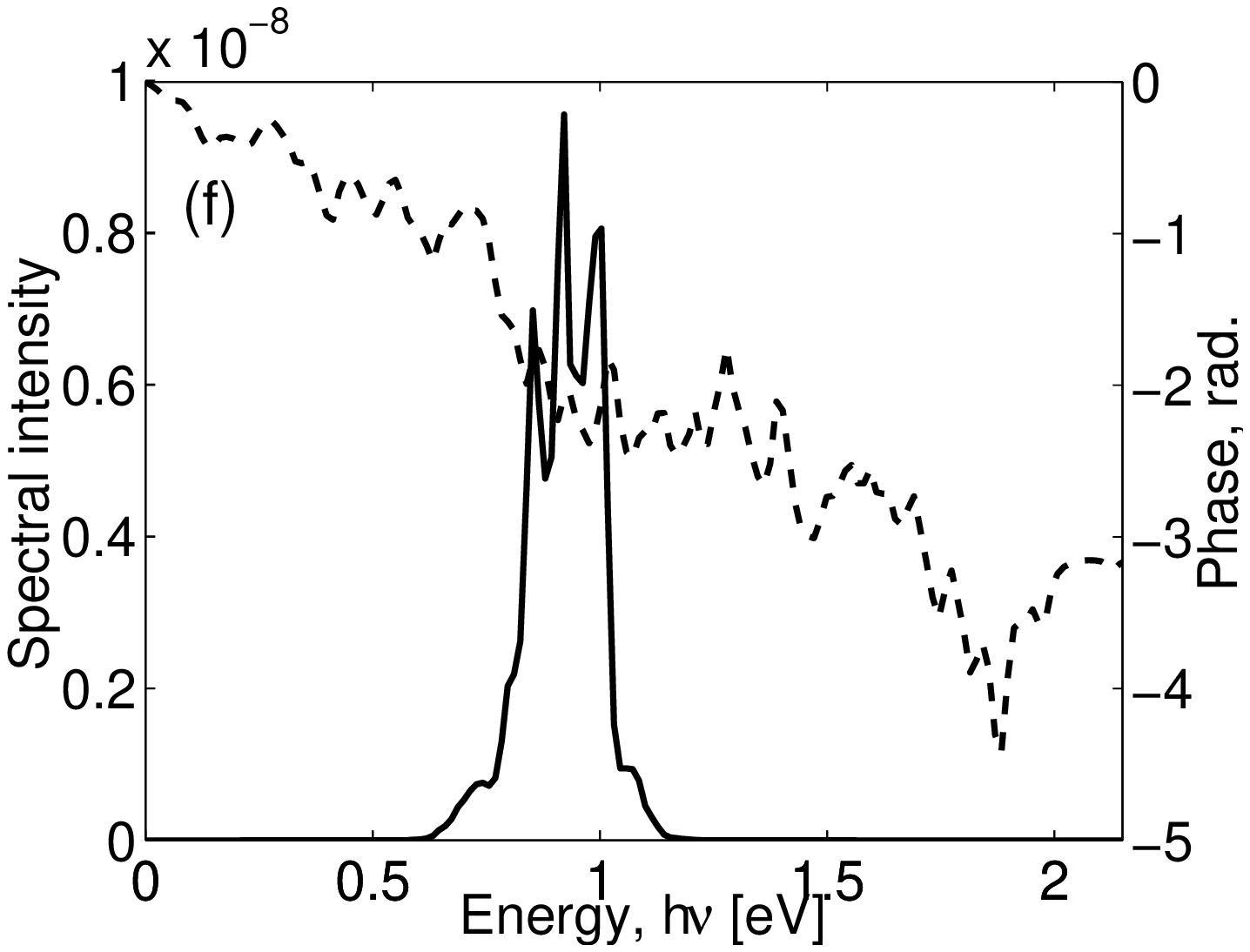}
\caption{\label{fig2} \footnotesize{ (a) Time dependence of a
Gaussian external field pulse with duration $10$ fs. (b) Spectral
intensity and phase of the incoming pulse shown in (a). Note, the
phase of the pulse is set to zero. (c) Time dependence of the local
field excited in the target region by the pulse (a). (d) Spectral
intensity and phase of the local field shown in (c). (f) Time
dependence of the external field that excites the total field in the
target region, same as in (a). (e) Spectral intensity and phase of
the incoming pulse shown in (f).}}
\end{figure}

Some examples are shown in Fig.~\ref{fig2}: A $10$~fs
transform-limited Gaussian pulse, Fig.~\ref{fig2}(a,b) incident on
the nanostructure will result in a local field shown in
Fig.~\ref{fig2}(c,d). Note the substantial difference in the shape
of the local field as compared to the incident field, particularly
the parasitic ``ringing'' following the main pulse. This is not
surprising, since in the simplest view the collective response of
the electrons in the nanostructure comprise a damped oscillator with
one or more eigenmodes, the frequencies and amplitudes of which are
determined by the geometry and electron concentration. The solution
to the inverse problem of obtaining the local field of the form
shown in Fig.~\ref{fig2}(a,b) is displayed in Fig.~\ref{fig2}(e,f)
and is easy to understand by noticing that the spectral phase of the
incident field is simply the inverted phase of the nanostructure
response. Similarly the spectral amplitude structure of the incident
field compensates for the spectral variations of the nanostructure
response to result in a smooth amplitude spectrum of the local
field. A consequence of this is that the ``ringing'' on the drive
pulse is exactly out of phase with the nanostructure ``ringing'', so
that the two contributions cancel each other.

In conclusion,  we have demonstrated theoretically the feasibility
of optimal coherent control near metallic nanostructures, and how to
use the resonant properties of the nanostructure to generate
strongly enhanced, spatially {\it and} temporally localized control
field. We have shown that the local field shape deformation and
parasitic ''ringing" can be successfully eliminated. For relatively
weak fields one can readily find an external control field, that for
the particular geometry and given carrier concentration in the
nanostructure will generate the necessary local control field at a
given point.  The solution of the inverse problem always exists and
does not depend on the shape of the optimal local pulse. In
practice, the {\it optimal} control pulse can be found using
standard closed loop techniques \cite{closed_loop}, because the
exact response of each given nanostructure will be nearly impossible
to measure.
%Significantly,
%however,
%the requirements to
%precompensate for the nanostructure response are not stringent, that
%is phase and spectral variations are mild and
%the broadband nature
%of the plasmon resonance peak ensures the possibility of broadband
%(i.e. femtosecond) excitation.
Moreover, we studied the frequency dependence of the local field at
various points within the nanostructure. Although the inhomogeneous
local field amplitude may vary significantly from point to point, we
discovered that the phase of the field is much less sensitive to
spatial variations. This provides additional tolerance to the
position fluctuations of the controlled molecule.

%Let us discuss some restrictions of our model.
In our simulations we used a model of non-interacting electrons
trapped in the nanostructure. Such simplifications will potentially
lead to underestimation of the Landau damping \cite{eguiliz},
resulting in narrower resonance peaks. However, the position of the
main resonance peaks obtained, for example, using the {\it ab
initio} Density Functional Theory, should be similar to our
simplified model \cite{eguiliz}.
 The applicability of the linear response theory, used in this study, is limited to
relatively weak fields. For many optimal control applications this
assumption holds. However, in some cases, the field intensity can be
high enough, resulting in significant temperature rise and strong
nonequilibrium electron distribution \cite{review}, which will
require the use of the nonlinear response theory. Further, the
breakdown of the linear response approximation will make the
response kernel ${\bf {\hat Z}}({\bf r}_0,\omega)$ dependent on the
amplitude of the external field. In this case one cannot guarantee
the inversion of Eq.(\ref{frequency2}) for an arbitrary optimal
local field.

 The electron-electron, electron-phonon
scattering and other processes, having unpredictable, stochastic
nature lead to thermalization of the electron gas in the
nanostructure and eventually to the loss of  coherence in the system
\cite{dephasing}. However, if the controlled molecule is not in the
direct contact with the nanostructure, these processes would not
limit the possibility of the optimal control. Because of the
relatively large number of electrons in the nanostructure, the local
electric field has no uncertainties related to the quantum dynamics
of electrons, and the nanostructure effectively acts as a {\it
linear filter}. In this case the coherent control is limited by the
intrinsic decoherence times of the controlled molecule. However, as
we mentioned above, these times may be decreased due to the
dynamical screening effects \cite{fluer}.

Since the temporal profile of the local control field keenly depends
on the position of the controlled molecule on the nanostructure, one
needs to address possible mechanisms and controllability of spatial
localization of the controlled molecules. It was theoretically
demonstrated that the plasmon-generated spatially inhomogeneous
field between metal nanoparticles can serve as a trap for molecules
\cite{trapping_b_spheres} without direct contact to the
nanoparticles, which otherwise would lead to strong decoherence. It
was also shown \cite{levi,trapping_b_spheres} that the field
enhancement is maximum in the space between the nanoparticles. Thus,
properly arranged arrays of metal nanoparticles may offer {\it both}
trapping and field enhancement capabilities.

This work was performed, in part, at the Center for Integrated
Nanotechnologies, a U.S. Department of Energy, Office of Basic Energy Sciences
user facility.  Los Alamos National Laboratory, an affirmative action equal
opportunity employer, is operated by Los Alamos National Security, LLC, for the
National Nuclear Security Administration of the U.S. Department of Energy under
contract DE-AC52-06NA25396.

\end{document}